          [2001/04/25 1.1 (PWD)]
\documentclass[letterpaper,twocolumn]{esapub} 

\usepackage{natbib}
\usepackage{graphicx}
\title{Gamma-Ray Line Observations with RHESSI}
\author{David M. Smith}
\affil{Department of Physics and Santa Cruz Institute for Particle Physics,
University of California, Santa Cruz, 1156 High Street, Santa Cruz, CA 95064 USA, dsmith@scipp.ucsc.edu}

\begin{document}


\maketitle

\begin{abstract}

The Reuven Ramaty High Energy Solar Spectroscopic Imager (RHESSI) has
been observing gamma-ray lines from the Sun and the Galaxy since its
launch in February 2002.  Here I summarize the status of RHESSI
observations of solar lines (nuclear de-excitation, neutron capture,
and positron annihilation), the lines of $^{26}$Al and $^{60}$Fe from the 
inner Galaxy, and the search for positron annihilation in novae.

\end{abstract}

\section{Introduction}

The \it Reuven Ramaty High Energy Solar Spectroscopic Imager
(RHESSI)\rm, which was launched on February~5, 2002, is part of NASA's Small
Explorer series of satellites \citep{lin}.  \it RHESSI \rm was
designed to perform imaging and spectroscopy of solar flares in the
hard x-ray and gamma-ray range, with both spatial and spectral
resolution far superior to previous missions.  It uses Rotating
Modulation Collimators (RMCs) for solar imaging down to
2.3''\citep{hur}.  Its nine coaxial germanium detectors cover
the range from 3~keV to 17~MeV with extremely good energy
resolution ($\rm{E}/\Delta\rm{E}$ around 400 from 1--2~MeV) 
\citep{sm1}.  The detectors are in the rear of the spacecraft;
the spacecraft bus is wrapped around the central tube that contains
the optical axis.  Each detector is segmented into a thin front
segment, which records the copious hard x-ray photons from a solar
flare, and a thick rear segment built to observe solar gamma-rays at
higher energies that penetrate the front segment.

Only {\it RHESSI's} coarsest grids are thick enough to effectively
modulate gamma rays above 1~MeV.  Imaging is therefore limited to
35'' at the 2.223~MeV line from neutron capture in solar flares, and
to 3' for the 4 and 6 MeV lines from carbon and oxygen de-excitation.
Solar spectroscopy can be performed across the entire energy range 
of the detectors, subject to the limitations of photon statistics
at gamma-ray energies.

Because these detectors have no shielding, and because the spacecraft
is very light, the {\it RHESSI} detectors can also see emission at
gamma-ray energies from any direction in the sky.  For photons on the
order of 1~MeV or higher, in fact, the instrument's effective area
(about 29 cm$^2$ at 1809 keV) has little dependence on the
direction of incidence, with the exception of a $\sim$30\%  depression 
at near-solar angles due to the effects of the grids, absorption in
the front segments, and opacity of parts of the spacecraft bus.
Since {\it RHESSI} is in low-Earth orbit, the
spectrum of emission from any given point in space can be derived by
using the Earth as an occulter, subtracting blocked from unblocked
data to determine background, if the source in question is the
dominant source in the sky at the wavelength (and during the time
period) under consideration.

\section{Solar Flare Observations}

Observations of gamma-ray lines due to nuclear interactions
in flares can help discriminate among models of ion acceleration.
The data and the models meet in the intermediate terrain of 
the energy spectra, composition, location, and directivity of
the accelerated ions: the gamma rays can be used to deduce these
distributions and the models make varied predictions.

Much has been learned from these lines using
scintillation detectors with moderate energy resolution, notably
the Gamma-Ray Spectrometer on the Solar Maximum Mission 
\citep[e.g.]{sm97}.  With the advent of {\it RHESSI}, 
gamma rays in flares have been imaged with a spatial resolution
comparable to the size of the emitting regions, and their spectra
have been observed with high energy resolution for the
first time (a feat that has
now also been performed by {\it INTEGRAL} \citep{gr04}).

A special volume of the {\it The Astrophysical Journal Letters}
presented a wide range of {\it RHESSI} results on the X4.8-class
flare of 23 July, 2002, a copious emitter of gamma rays.  Figure~1
shows the lightcurve of the flare in several energy bands.  Figure~2
shows an overview of the spectrum across {\it RHESSI's} range
of coverage.  The results reported in that volume
are summarized in \citet{lin03}, and include:

\begin{itemize}
\item The discovery \citep{hu03} that the 2.223~MeV neutron-capture line in this flare,
a good tracer of the interactions of accelerated ions, was centered
at a position significantly offset from the distribution of hard x-rays
from electron bremsstrahlung.  The spatial separation between accelerated
ions and electrons in this flare puts strong constraints on acceleration
models.
\item 
The unexpectedly large width (8.1 $\pm$ 1.1)~keV
of the positron-annihilation line \citep{share03}.  Two possible local
conditions could prevail to give a line this broad: a very high-temperature
yet dense and extended plasma of (4--7)$\times 10^{5}$K, or a similarly
extensive plasma at a temperature restricted to a very narrow range around
6000~K.  Neither environment is expected to be typical of flaring atmospheres,
but little is known about their true conditions.
\item
Significant redshifts and broadening found in the nuclear de-excitation lines 
of C, O, Ne, Mg, Si, and Fe \citep{sm03s}.  These Doppler effects are
due to gamma emission while the nucleus is still recoiling from a collision,
and have been observed before \citep{sm97}, but not with high energy resolution.
The redshifts were unexpectedly high for a flare on the solar limb, suggesting
either a magnetic loop tilted toward the observer or an extreme beaming
of the ions toward the solar surface (but see below).
\item
A very broad shape for the lines from interactions of accelerated $\alpha$
particles with ambient $^4$He \citep{share2}.  These lines (from $^7$Be 
at 429~keV
and $^7$Li at 478~keV) combine to make a single broad peak of characteristic
shape, indicating that the angular distribution of the accelerated particles
was {\bf not} beamed.
\item
A study, using an advanced model of ion propagation in magnetic loops, of the
time delay of the neutron capture line compared to prompt nuclear emissions
\citep{mur03}.  The conclusion was that pitch-angle scattering of the ions
must be taking place in the coronal part of the loop.
\end{itemize}

Work currently in progress includes a study \citep{shih04} of variability of 
line ratios between the two halves of this flare (marked in Figure~1), and
extensive studies of the gamma-ray emission in the large X-class flares
of October/November 2003.

\begin{figure}
\centering
\includegraphics[width=\linewidth]{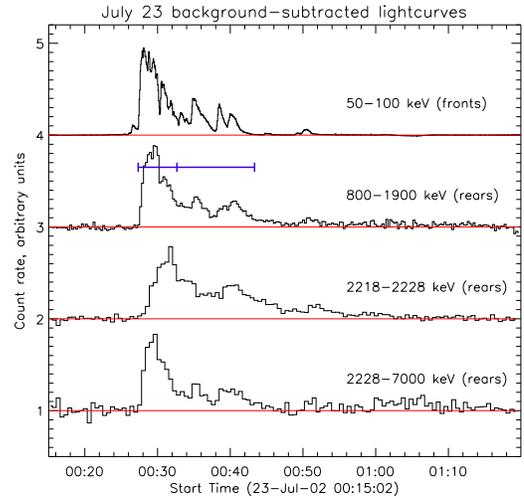}
\caption{Background-subtracted {\it RHESSI} lightcurves of the 
X4.8 flare of 23 July, 2002.  The top plot shows the hard x-rays
from electron bremsstrahlung, and the other three channels are
dominated by gamma rays from nuclear interactions of accelerated
ions.  The third plot is the lightcurve of the narrow line from
neutron capture at 2.223~MeV, showing the delay due to the neutron
thermalization time.  
}
\end{figure}

\begin{figure}
\centering
\includegraphics[width=\linewidth]{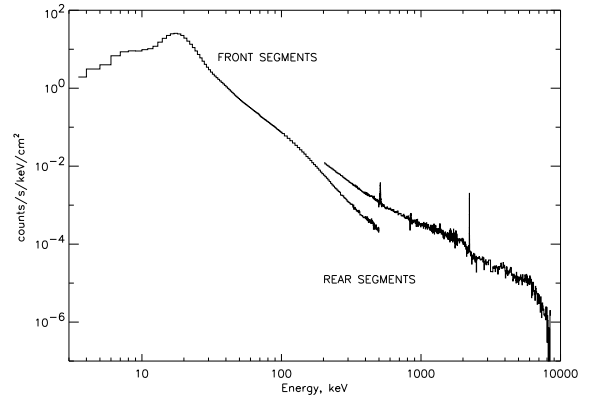}
\caption{The broad-band count spectrum of the flare of 23 July, 2002
as observed with {\it RHESSI}, demonstrating the extremely large 
dynamic range of the instrument in both energy and flare flux
({\it RHESSI} has also usefully studied microflares over five orders
of magnitude smaller than this event).  The positron-annihilation
line at 511~keV and the neutron capture line at 2.223~MeV are clearly visible,
as is the cutoff above 7~MeV where there is no longer a strong contribution
from nuclear lines.  The decline below 20~keV is due to aluminum
attenuators that are moved in front of the detectors to control
the count rate in bright flares.
}
\end{figure}

\section{Galactic Nucleosynthesis}

\subsection{$^{26}$Al}

\begin{figure}
\centering
\includegraphics[width=\linewidth]{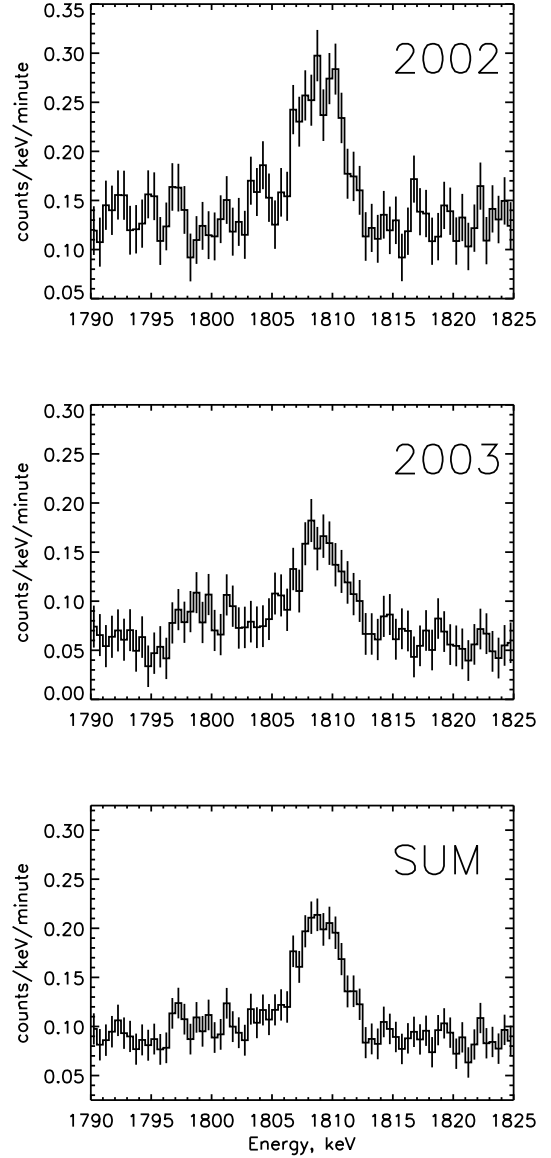}
\caption{1809 keV line profiles for the early mission
(2002 March--November), the later mission 
(2002 December -- 2003 November) and
the sum of the two periods.  Radiation damage in the
detectors adds a tail to the line shape and degrades
the photopeak response, and this effect increases
with time.
}
\end{figure}

\begin{figure}
\centering
\includegraphics[width=\linewidth]{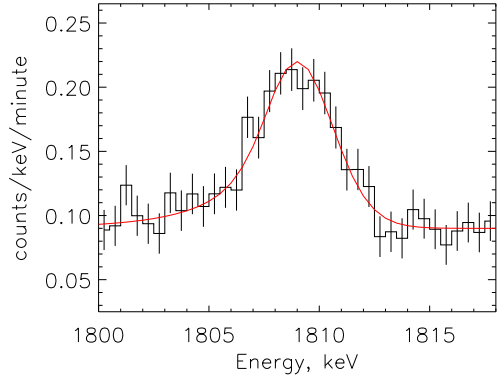}
\caption{The 1809 keV profile for the whole mission with
the instrumental lineshape superimposed.  It is clear that
there is no signficant broadening beyond the instrumental
response.
}
\end{figure}

The radioactive lines from $^{26}$Al and $^{60}$Fe were extracted
for the inner Galaxy, a region
defined as covering $\pm$ 30$^{\rm{\circ}}$ in Galactic longitude and
$\pm$ 5$^{\rm{\circ}}$ in Galactic latitude.  When \it none \rm of
this region was occulted by the Earth, each minute of data was summed
into the total "source" spectrum, and when \it all \rm of it was
occulted, the spectrum was used to create a database of background
spectra.  Similar background subtraction techniques have been
used for observations with the 
\it Solar Maximum Mission \rm Gamma-Ray Spectrometer ({\it SMM}/GRS)
\citep{ha1,ha2} and the Burst And Transient Source Experiment on the
\it Compton Gamma-Ray Observatory \rm \citep{sm3,sm4}.

Although the fluxes thus derived have virtually no dependence on the
distribution of the isotopes {\it within} the inner Galaxy box, the
distribution  {\it outside} will affect the recorded flux in a complicated
way, depending on whether it resides in parts of the sky more often
appearing in source or background spectra.  This issue is being studied
in detail by \citet{wu04}.

I reported the $^{26}$Al flux for the 
first nine months of the {\it RHESSI} mission
to be $(5.71 \pm 0.54) \times 10^{-4}$ ph~cm$^{-2}$~s$^{-1}$
\citep{sm2,sm04},
comparable to the value found by the GRIS balloon \citep{naya} but
significantly higher than the sum of the COMPTEL map in this range of
Galactic longitude \citep{diehl}.  I have recently discovered a
significant error in the calculation of the instrumental effective
area, however.  The correct effective area, taking a weighted average
over all the angles of incidence during the different pointing periods
used, should have been 29.3~cm$^2$ rather than the 20.5~cm$^2$ quoted
earlier.  The initial result should then have been $(3.99 \pm 0.38)
\times 10^{-4}$ ph~cm$^{-2}$~s$^{-1}$, in much better agreement with
the results from COMPTEL and other instruments.  The results below and
in the future will use the correct instrumental response.  With minor
improvements to the data selection and background selection algorithms
since \citet{sm2}, I now find $(3.89 \pm 0.36) \times 10^{-4}$
ph~cm$^{-2}$~s$^{-1}$ for the same data set; the change other than that
due to the instrument effective area is negligible.

The later data set (Figure~3, middle panel) gives a flux of only
$(3.03 \pm 0.36) \times 10^{-4}$ ph~cm$^{-2}$~s$^{-1}$ when fit, as
the first data set was, with a simple Gaussian.  It is clear, however,
from the shape of the line, that there is a significant tailing of
the line shape due to radiation damage in the detectors as the mission progresses.
This tailing is more obvious and more quantifiable in the background lines,
which can be accumulated with extremely small error bars over long time
intervals.  By studying these lines, particularly the line of $^{40}$K
(which will have a constant flux throughout the mission since it is a
natural radioactivity), we can determine the effect that radiation damage
has on the measured flux for a Gaussian fit.  This correction, based on 
the data, has also been checked with simulations
(these simulations were made before the start of the mission and have
accurately predicted the degree of radiation damage versus time).
The corrected flux for the later data set is
$(3.46 \pm 0.41 \pm 0.18) \times 10^{-4}$ ph~cm$^{-2}$~s$^{-1}$, where
the second error listed is a generous margin for systematic errors in the 
radiation-damage correction.  The corrected value is compatible
within statistical errors with the measurement from the first year.

The summed data set for the entire mission
to date gives $(3.35 \pm 0.24) \times 10^{-4}$ ph~cm$^{-2}$~s$^{-1}$
uncorrected, and $(3.69 \pm 0.27 \pm 0.11) \times 10^{-4}$ ph~cm$^{-2}$~s$^{-1}$
corrected for radiation damage.  The effective area for the full data set
is slightly lower (28.6~cm$^2$), since it includes a December/January period when the
Galactic Center is near the Sun.  Now that the effective area has been
corrected, the {\it RHESSI} data no longer support the tentative conclusion 
\citep{Naya98}, based on the
GRIS balloon data, that there might be a low-surface-brightness,
highly diffuse component of the 1809~keV line not visible to imaging
instruments like COMPTEL.  Rather, they are now consistent with the results
from COMPTEL and {\it SMM}/GRS \citet{ha1}, another wide-field instrument
like GRIS and {\it RHESSI}.

The GRIS result also included a measurement of the intrinsic line width
(i.e. with instrumental broadening removed) that was surprisingly
broad (5.4 (+1.4, -1.3)~keV) \citep{naya}, which was difficult to
understand theoretically, since supernova ejecta come to rest in the
interstellar medium on a timescale much shorter than the halflife of
$^{26}$Al.  The {\it RHESSI} result for the intrinsic width in the
first nine months of the mission was 2.03~(+0.78, -1.21)~keV, in clear
disagreement with GRIS and consistent with the $\sim$1~keV width
expected from Galactic rotation \citep{kret}.  Including the data
through 2003 November, I find 0.9~(+1.1, -0.9)~keV.  Figure~4 shows
the line shape (also shown in the bottom panel of Figure~3, with the
instrumental line shape superimposed.  The instrumental shape, which
has a full width at half maximum of 4.5~keV and a significant
low-energy tail, was derived from background lines at nearby energies.

\subsection{$^{60}$Fe}

The radioactivity of Galactic $^{60}$Co, the daughter of $^{60}$Fe, 
which is expected to be produced
in core-collapse supernovae that may also produce much of the Galactic 
$^{26}$Al, has not previously been observed.  Lines are expected at
1173~keV and 1332~keV with equal intensity.  Models
of nucleosynthesis have predicted that the ratio of the
$^{60}$Fe/$^{26}$Al lines from the population of Galactic
supernovae is on the order of 15\% \citep{timmes}, but higher
ratios have also been calculated more recently
\citep{rau,lim}, leading to a renewal of the suggestion that supernovae
may not be the dominant source of $^{26}$Al \citep{prant}.

An analysis of the first 14 months of {\it RHESSI} data \citep{sm04},
which combined the two lines to increase the statistical significance
of the result, found a flux per line of ($16 \pm 5$)\% of the 1809~keV
flux.  This analysis was similar to that performed for the 1809~keV
line, except that data shortly after passes through the South Atlantic
Anomaly (SAA) were removed due to the presence of an SAA-activated
line, probably from $^{60}$Cu, at 1332 keV.  Summing the data through
November 2003, using an improved algorithm to better identify
SAA-contaminated periods, and using corrected instrumental 
effective areas at both energies,
the ratio is now found to be ($9.7 \pm 3.9$)\%.
Most of the difference is due to the incorporation of new data rather
than the changes in the analysis method (the error in effective
areas was comparable at both energies).  The radiation-damage
correction is significantly less important at these energies than at
1809~keV and has not been applied.  Figure~5 shows the individual
lines and their sum.  The 1173~keV line has a significance of only
1.4$\sigma$, with a flux of $(2.8 \pm 2.0) \times 10^{-5}$
ph~cm$^{-2}$~s$^{-1}$, and the 1332~keV line has a flux of $(4.2 \pm
1.7) \times 10^{-5}$ ph~cm$^{-2}$~s$^{-1}$ (2.4$\sigma$), giving an
average of $(3.6 \pm 1.4) \times 10^{-5}$ ph~cm$^{-2}$~s$^{-1}$
(2.6$\sigma$).  Further data will be added as the mission
progresses.

\begin{figure}
\centering
\includegraphics[width=\linewidth]{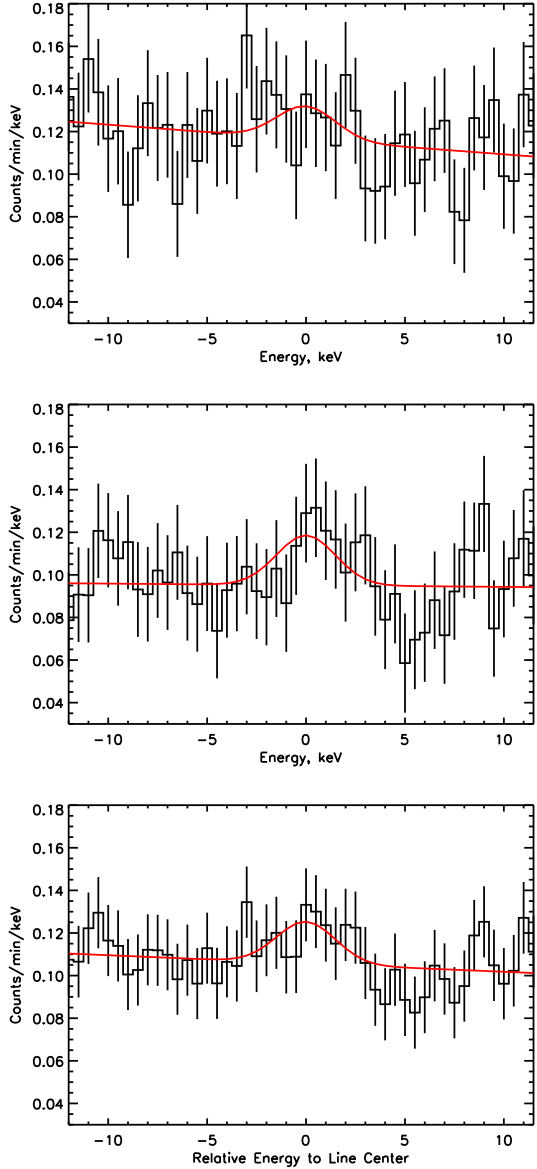}
\caption{The 1173~keV (top) and 1332~keV (middle) profiles 
for the whole mission,
and their sum (bottom), shifted to put the expected
line center at zero energy.}
\end{figure}

\subsection{Positron Annihilation in Novae}

Models of classical novae from either CO or ONe white dwarfs predict
copious emission in the positron annhilation line at 511 keV from
radioactive decay of $^{13}$N and $^{18}$F, starting at the moment of
explosion \citep{hern}.  This emission occurs long before the nova
could be detected in visible light, so an instrument that views a
substantial portion of the sky is required to detect this emission.
So far, only upper limits have been obtained, from BATSE on {\it CGRO}
\citep{hern00} and TGRS on {\it Wind} \citep{harris1, harris2}.  Both
studies obtained upper limits on the annihilation line from $^{18}$F,
which has a duration on the order of half a day.  The 3$\sigma$ upper
limits on individual novae averaged $2.4 \times 10^{-3}$
ph~cm$^{-2}$~s$^{-1}$ over 6-hour intervals with TGRS \citep{harris1}
and $2.1 \times 10^{-3}$ ph~cm$^{-2}$~s$^{-1}$ over 12-hour intervals
with BATSE.  Despite the TGRS detector having only a tiny fraction of
BATSE's area, it was able to reach comparable sensitivity due to its
high energy resolution, since the high velocity of the nova ejecta
would give a blueshift to the material on the near side of the white
dwarf, moving the line a few keV away from the bright background line.
{\it RHESSI} has comparable energy resolution and nearly an order of
magnitude more detector volume than TGRS, but its background level is
about a factor of two higher due to the proximity of the Earth.  Thus
we expect {\it RHESSI} to have between a factor of 2 and 2.5 better
sensitivity for this measurement than TGRS and BATSE.

Since those limits were published, better laboratory
data and modeling have resulted in new predictions that the line 
from $^{18}$F will be considerably lower than previously thought,
putting it beyond the sensitivity of {\it RHESSI} and requiring
either a very serendipitous detection by {\it INTEGRAL}/SPI in its
field of view, or else the launch of
a future, highly sensitive mission with full-sky or nearly full-sky
coverage such as NASA's planned Black Hole Finder Probe and 
Advanced Compton Telescope (ACT).  These newer estimates
\citep{hern,kud} also predict, however, that the line from $^{13}$N,
lasting only for the first couple of hours of the explosion,
could be much brighter, particularly for CO novae, and could
therefore present a good target for {\it RHESSI}; it is also
expected, by virtue of appearing earlier, to have a higher redshift.

Although there has not been any particularly close nova since {\it
RHESSI's} launch, the month of September 2002 provided two distant
novae in the direction of Sagittarius \citep{liller,haseda}.  Figure~6
shows {\it RHESSI} spectra in two hour accumulations from September
1--24, a period that almost certainly contains the outbursts from
both novae.  Background has been subtracted by using data 15 orbits
(almost exactly one day) ahead and behind the interval of interest.
This provides good subtraction to a level of a few percent, but the
Figure shows that even this results in very large residuals at the
unshifted annihilation line.  Superimposed on the data in Figure~6 is
{\it RHESSI's} response to a blueshifted annihilation line at a flux
level of 0.01 photons cm$^{-2}$s$^{-1}$, equivalent to some of the
brighter predictions \citep{hern,kud} for a distance of 1~kpc; this is
intended more to give a notion of what could be observed from a closer
event than to represent an expected signal from these particular
novae.

\begin{figure}
\centering
\includegraphics[width=\linewidth]{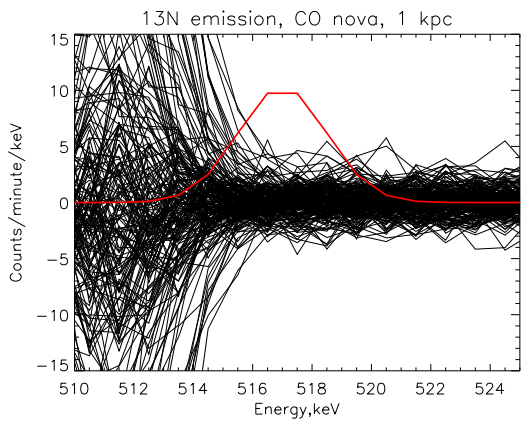}
\caption{Background-subtracted 2-hour count spectra (see text)
for 1--24 September 2002.  Note the large background variability
at the unshifted 511~keV line.  For comparison, a blueshifted
line is shown at 0.01 photons cm$^{-2}$s$^{-1}$, an optimistic
estimate for $^{13}$N production in a CO nova at 1~kpc.
}
\end{figure}

\section{Future Work}

All of the studies described here are ongoing.  We are currently studying
the lines from the flares of October/November 2003 and making plans
for other solar studies, including a search for steady
ion acceleration by studying the quiet Sun for low-level line emission and
a search for radioactive $^{56}$Co in the aftermath of large flares
\citep{ramaty}.  The study of the $^{26}$Al line is bringing us together
with {\it INTEGRAL}/SPI to combine the information at different spatial
scales \citep{wu04}, and the $^{60}$Fe result is simply awaiting the
gradual accumulation of further data.  We will survey the entire sky
for flashes of annihilation radiation (from classical novae or
any other source), as has been done with other instruments
\citep{harris2,sm3,cheng}.  Finally, we will use the
same technique that was used for the Galactic center lines to see if
we can obtain a high-resolution measurment of the 1157~keV line from
$^{44}$Ti in the Cas~A supernova remnant.

\section*{Acknowledgments}

This work is supported by NASA contract NAS5-98033.


\begin{thebibliography}{}

\bibitem[Cheng et al.(1998)]{cheng}
Cheng, L. et al. (1998), ApJ, 503, 809

\bibitem[Diehl et al.(1995)]{diehl}
Diehl, R. et al. (1995), A\&A, 298, 445

\bibitem[Gros et al.(2004)]{gr04}
Gros, M. et al. (2004), this volume

\bibitem[Harris et al.(1990)]{ha1}
Harris, M. J. et al. (1990),  ApJ, 362, 135

\bibitem[Harris, Share, \& Leising(1994)]{ha2}
Harris, M. J., Share, G. H., \& Leising, M. D. (1994), ApJ, 433, 87

\bibitem[Harris et al.(1999)]{harris1}
Harris, M. J. et al. (1999),  ApJ, 522, 424

\bibitem[Harris et al.(2000)]{harris2}
Harris, M. J. et al. (2000),  ApJ, 542, 1057

\bibitem[Haseda et al.(2002)]{haseda}
Haseda, K. et al. (2002), IAUC 7975

\bibitem[Hernanz et al.(2000)]{hern00}
Hernanz, M. J. et al. (2000), {\it Proceedings of the 5th
Compton Symposium}, AIP Conf. Proc. 510, 82

\bibitem[Hernanz \& Jos{\'{e}}(2004)]{hern}
Hernanz, M. J. \& Jos{\'{e}}, J. (2004), New Ast. Rev., 48, 35

\bibitem[Hurford et al.(2002)]{hur}
Hurford, G. J. et al. (2002), \it Solar Physics\rm, {\bf 210},~61

\bibitem[Hurford et al.(2003)]{hu03}
Hurford, G. J. et al. (2003), ApJL, 595, L77

\bibitem[Kretschmer et al.(2003)]{kret}
Kretschmer, K., Diehl, R., \& Hartmann, D. H. (2003), A\&A, 412, L47

\bibitem[Kudryashov et al.(2000)]{kud}
Kudryashov, A. D., Chuga{\u{i}}, N. N., \& Tutukov, A. V. (2000),
Ast. Rep. 44, 170

\bibitem[Liller et al.(2002)]{liller}
Liller, W. et al. (2002), IAUC 7971

\bibitem[Limongi \& Chieffi(2003)]{lim}
Limongi, M. \& Chieffi, A. (2003), ApJ, 592, 404

\bibitem[Lin et al.(2002)]{lin}
Lin, R. P. et al. (2002), {\it Solar Physics}, {\bf 210},~3

\bibitem[Lin et al.(2003)]{lin03}
Lin, R. P. et al. (2003), ApJL, 595, L69

\bibitem[Murphy et al.(2003)]{mur03}
Murphy, R. J. et al. (2003), ApJL, 595, L93

\bibitem[Naya et al.(1996)]{naya}
Naya, J. E. et al. (1996),  Nature, 384, 44

\bibitem[Naya et al.(1998)]{Naya98}
Naya, J. E. et al. (1998),  ApJL,  499, L169

\bibitem[Prantzos(2004)]{prant}
Prantzos, N. (2004), astro-ph/0402198, accepted in A\&A 

\bibitem[Ramaty \& Mandzhavidze(2000)]{ramaty}
Ramaty, R. \& Mandzhavidze, N. (2000), {\it Highly Energetic Physical 
Processes and Mechanisms for Emission from Astrophysical Plasmas},
Proc. IAU Symp. \#195, 123

\bibitem[Rauscher et al.(2002)]{rau}
Rauscher, T. et al. (2002), ApJ, 576, 323

\bibitem[Share et al.(2003a)]{share03}
Share, G. H. et al. (2003), ApJL, 595, L85

\bibitem[Share et al.(2003b)]{share2}
Share, G. H. et al. (2003), ApJL, 595, L89

\bibitem[Share \& Murphy(1997)]{sm97}
Share, G. H. \& Murphy, R. J. 1997, ApJ, 485, 409

\bibitem[Shih et al.(2004)]{shih04}
Shih, A. Y. et al. (2004), in preparation

\bibitem[Smith et al.(1996a)]{sm3}
Smith, D. M., et al. (1996a), ApJ, 471, 783

\bibitem[Smith et al.(1996b)]{sm4}
Smith, D. M., et al. (1996b), A\&AS, 120, 361

\bibitem[Smith et al.(2002)]{sm1}
Smith, D. M. et al. (2002), \it Solar Physics\rm, {\bf 210},~33

\bibitem[Smith et al.(2003a)]{sm03s}
Smith, D. M. et al. (2003), ApJL 595, L81

\bibitem[Smith(2003)]{sm2}
Smith, D. M. (2003), ApJL, 589, L55

\bibitem[Smith(2004)]{sm04}
Smith, D. M. (2004), New Ast. Rev., 48, 89

\bibitem[Timmes et al.(1995)]{timmes}
Timmes, F. X., et al. (1995), ApJ, 449, 204

\bibitem[Wunderer et al.(2004)]{wu04}
Wunderer, C. et al. (2004), this volume


\end{thebibliography}

\end{document}